\documentclass[sigconf,nonacm]{acmart}

\setcopyright{none}
\acmDOI{}
\acmISBN{}
\acmConference{}{}{}
\acmBooktitle{}
\AtBeginDocument{%
  }

\usepackage{booktabs}
\usepackage{multirow}
\usepackage[table]{xcolor}
\usepackage{graphicx}
\usepackage{array}
\usepackage{makecell}
\usepackage{pifont}

\usepackage{subcaption}

\newcommand{\cmark}{\ding{51}}
\newcommand{\xmark}{\ding{55}}
\newcolumntype{L}[1]{>{\raggedright\arraybackslash}p{#1}}
\usepackage{CJKutf8}

\usepackage{tabularx}

\newcolumntype{Y}{>{\centering\arraybackslash}X}

\newcommand{\framework}{\textsc{FinNextAssist}}

\makeatletter
\renewcommand{\@authorfont}{\LARGE}
\renewcommand{\@affiliationfont}{\large}
\@ifundefined{hyxmp@parse@acmart}{}{\let\hyxmp@parse@acmart\relax}
\makeatother

\begin{document}

\title{\raisebox{-0.5em}{\includegraphics[height=1.5em]{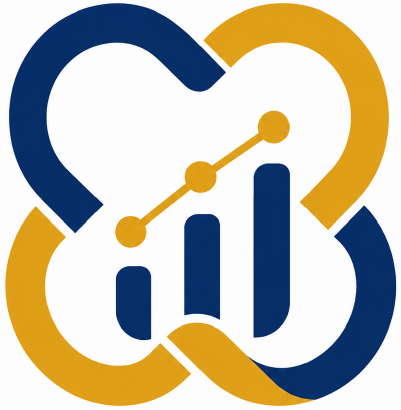}} \framework{}: Towards Professional Financial Deep Research Assistant}

\author{%
Xiangyu Li$^{1}$, Fengbin Zhu$^{2,*}$, Xuan Yao$^{2}$, Siyu Liu$^{3}$, Xiaoluan Liu$^{4}$, Chao Wang$^{5}$,\\
Huanbo Luan$^{5}$, Xiaofen Xing$^{1}$, Xiangmin Xu$^{1}$, Ke-Wei Huang$^{2}$, Richang Hong$^{6}$, Tat-Seng Chua$^{2}$
}
\affiliation{%
  \institution{%
    $^{1}$South China University of Technology;
    $^{2}$National University of Singapore;
    $^{3}$iFLYTEK CO., LTD;
    $^{4}$Central University of Finance and Economics;
    $^{5}$6Estates;
    $^{6}$Hefei University of Technology
  }
  \country{%
    65603605lxy@gmail.com, fengbin@nus.edu.sg, yaoxuan@nus.edu.sg,
    siyuliu114@gmail.com, xiaoluanliu@email.cufe.edu.cn, wangchao@6estates.com,
    luanhuanbo@6estates.com, xfxing@scut.edu.cn, xmxu@scut.edu.cn,
    dishkw@nus.edu.sg, hongrc.hfut@gmail.com, dcscts@nus.edu.sg
  }
}
\email{}
\renewcommand{\shortauthors}{Xiangyu Li et al.}

\begin{abstract}
  Deep Research (DR) agents have demonstrated strong capabilities in complex, research-oriented tasks through autonomous planning, iterative retrieval, multi-step reasoning, and structured reporting. 
  However, adapting DR agents to finance introduces unique challenges: financial analysis demands the joint completion of heterogeneous sub-tasks spanning diverse data types, tools, and analytical workflows. We identify three key requirements for a professional financial DR agent: integration of authoritative, heterogeneous financial data sources; specialized analytical tools and skills; and dedicated sub-agents for domain-specific sub-tasks. Building on these principles, we propose \textbf{\framework{}}, an end-to-end deep research framework designed for professional financial analysis. \framework{} decomposes the research process into four stages: Task Planner, Evidence Compiler, Reasoning Engine, and Report Assembler,  and introduces two novel lightweight sub-agents: TabAgent, for cross-market financial table understanding, and HeteroAgent, for cross-modality heterogeneous financial data interpretation. 
  Extensive experiments on FinDeepResearch, the Finance Agent Benchmark, and FinTMMBench-Web show that \framework{} substantially outperforms both strong proprietary and open-source DR agents, with ablation studies confirming the contribution of each component across diverse markets and languages.

\end{abstract}


\maketitle
\hypersetup{%
  pdftitle={FinNextAssist: Towards Professional Financial Deep Research Assistant},
  pdfauthor={Xiangyu Li, Fengbin Zhu, Xuan Yao, Siyu Liu, Xiaoluan Liu, Chao Wang, Huanbo Luan, Xiaofen Xing, Xiangmin Xu, Ke-Wei Huang, Richang Hong, Tat-Seng Chua},
  pdfpublisher={},
  pdfcontactcountry={},
  pdfcontactemail={65603605lxy@gmail.com, fengbin@nus.edu.sg, yaoxuan@nus.edu.sg, siyuliu114@gmail.com, xiaoluanliu@email.cufe.edu.cn, wangchao@6estates.com, luanhuanbo@6estates.com, xfxing@scut.edu.cn, xmxu@scut.edu.cn, dishkw@nus.edu.sg, hongrc.hfut@gmail.com, dcscts@nus.edu.sg}
}

\begingroup
\renewcommand{\thefootnote}{\fnsymbol{footnote}}
\footnotetext[1]{Corresponding author: Fengbin Zhu (\href{mailto:fengbin@nus.edu.sg}{fengbin@nus.edu.sg}).}
\endgroup

\section{Introduction}
\label{sec:intro}

Deep Research (DR) agents, built upon and beyond large language models (LLMs), are capable of completing complex, research-oriented tasks through autonomous planning, iterative retrieval and reasoning, tool calling, and structured reporting~\cite{xi2025searchsurvey,li2025drsurvey,shi2025drsystematic}.
The rapid advancement of reasoning-capable large language models~\cite{deepseekai2025r1,openai2024o1,yang2025qwen3} and tool-augmented agentic frameworks~\cite{yao2023react} has provided the necessary foundations for their development, driving swift deployment across both proprietary systems~\cite{openai2025dr,google2024geminidr} and open-source projects~\cite{li2025webthinker,zheng2025deepresearcher}. 
Beyond general-purpose research assistance, an increasing number of works have begun adapting DR agents to high-stakes domains such as finance~\cite{jin2025finsight,zhu2025findeepresearch,liu2026towards}, highlighting both the paradigm's versatility and the pressing demand for specialized, trustworthy DR systems capable of operating reliably in domain-specific settings.

\begin{figure}[t]
\centering
\includegraphics[width=\columnwidth]{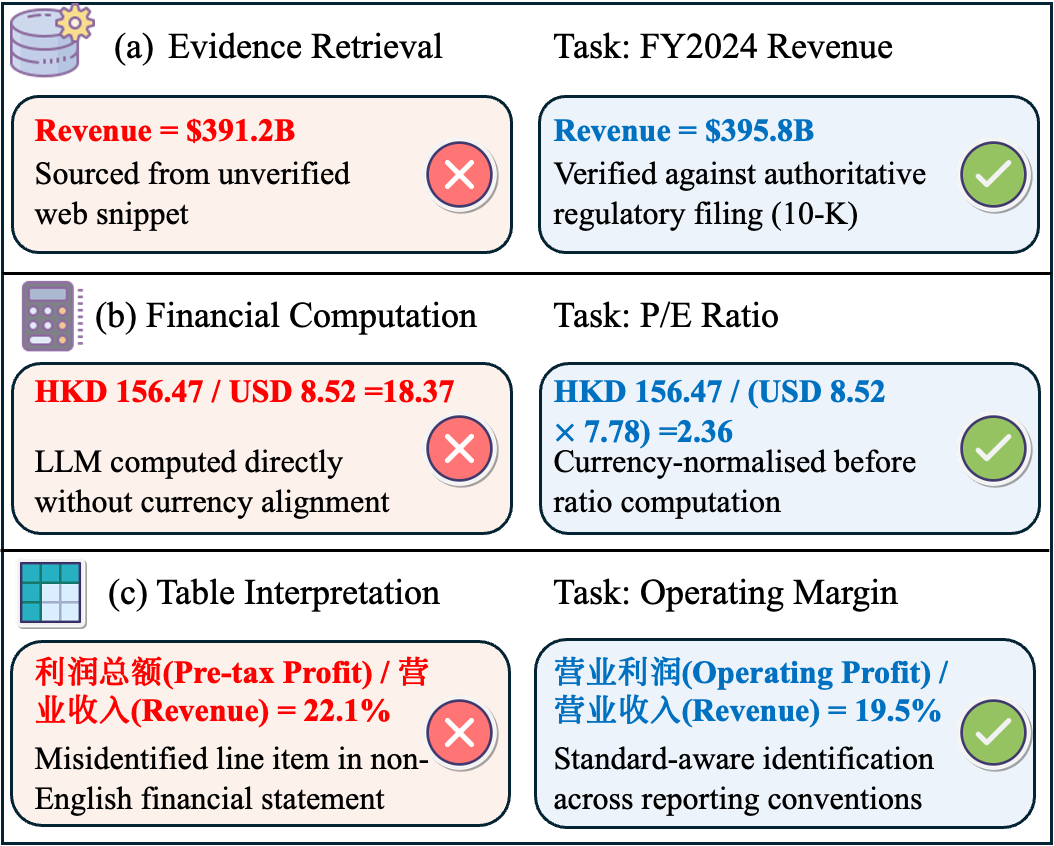}
\caption{Representative failures of existing Deep Research agents in financial analysis. }
\label{fig:intro}
\vspace{-10pt}
\end{figure}

However, adapting DR systems to the finance domain introduces unique and non-trivial challenges.
As demonstrated in FinDeepResearch benchmark~\cite{zhu2025findeepresearch}, a comprehensive financial analysis demands a DR agent to perform 
a diverse set of sub-tasks jointly: retrieval of regulatory filings, interpretation of financial tables, derivation of financial ratios, analysis of stock price dynamics, and interpretation of event-driven market reactions, among others.
These sub-tasks differ not only in the input content but also in the form of correctness they require: ratio computation demands exact numerical consistency; event-reaction analysis depends on precise temporal alignment; and qualitative assessment requires grounded semantic synthesis.
Nevertheless, in standard DR pipelines, numerical tables, unstructured textual disclosures, derived quantitative indicators, time-series market data, and narrative evidence are indiscriminately concatenated and flattened into a unified context. 
This forces LLMs to implicitly infer type boundaries and verification rules, thereby conflating heterogeneous reasoning requirements and substantially elevating the risk of analytical errors.

To effectively complete these diverse sub-tasks, we argue that an ideal financial DR agent must address three key requirements:
1) \textbf{Authoritative heterogeneous financial data sources}. Most existing DR systems~\cite{openai2025dr,gemini25DR}  rely on general-purpose search engines (e.g., Bing and Google Search) while overlooking the integration of domain-specific financial data spanning both unstructured sources (e.g., regulatory filings, financial news) and structured sources (e.g., market data, financial statements). As illustrated in Figure \ref{fig:intro} (a), errors can arise when web-retrieved snippets contain incorrect financial figures (e.g., revenue), whereas the values in authoritative regulatory filings are accurate.
2) \textbf{Specialized financial tools and skills}. Existing DR agents usually lack domain-aware tools essential for finance, such as currency conversion and standardized unit normalization. 
As demonstrated in Figure~\ref{fig:intro} (b), failure to normalize currencies before computing metrics such as the P/E ratio leads directly to erroneous results.
3) \textbf{Dedicated sub-agents for domain-specific sub-tasks}. For example, as shown in Figure~\ref{fig:intro} (c), the precise interpretation of financial tables poses a distinct reasoning challenge, as balance sheets, income statements, and cash flow reports are dense with domain-specific metrics and ratios. 

To instantiate this principle, we propose \textbf{\framework{}}, an end-to-end deep research framework for professional financial analysis. 
\framework{} decomposes the research process into four stages: 
1) \emph{Task Planner} analyzes the research query and target company context, decomposing the task into analytical dimensions with typed constraints, required evidence, data sources, and correctness predicates. 
2) \emph{Evidence Compiler} obtains raw evidence from domain-specific, heterogeneous public sources, including regulatory filings, financial data APIs, and web search results, and normalizes them into typed evidence bundles. 
3) \emph{Reasoning Engine} infers the analytical outcomes under the joint constraints of the task plan and typed evidence. It is equipped with two lightweight sub-agents: \textbf{TabAgent}, which specializes in cross-market financial table understanding, and \textbf{HeteroAgent}, which is designed specifically for cross-modality, heterogeneous financial data interpretation. 
Moreover, specialized financial skills and tools are applied to handle precise numerical operations, such as ratio computation, return calculation, and currency conversion, separately from the LLM generation path, thereby ensuring computational correctness.
4) \textit{Report Assembler} performs schema-conditioned synthesis, verifies structural invariants, and ensures that each output item is supported by a complete provenance chain.

We evaluate \framework{} on three financial research benchmarks, where it outperforms both strong proprietary DR agents and open-source DR systems.
On FinDeepResearch~\cite{zhu2025findeepresearch} it achieves an overall score of 48.5, improving over the strongest baseline by 10.5 points, with the largest gains in the Financial Performance and Market Performance sections and in the Recognition and Calculation capabilities, precisely the settings that demand reliable comprehension of structured statements, precise numerical computation, cross-source reconciliation, and temporal alignment between market events and price reactions.
It also ranks first in all eight evaluated markets, covering English, Simplified Chinese, Traditional Chinese, and Bahasa Indonesia, showing that the gains are robust across heterogeneous markets and languages.
On the Finance Agent Benchmark~\cite{bigeard2025finance}, \framework{} reaches 77.39\% accuracy, 20.68 points above the strongest baseline, and on FinTMMBench-Web~\cite{zhu2025fintmmbench} it exceeds the best Web-based baseline by 25.87 EM.
Ablation studies show that TabAgent contributes to table-centric extraction, line-item normalization, and metric computation while HeteroAgent enhances the interpretation of filings, market data, news events, and temporal dynamics, and backbone ablations confirm that the approach generalizes across LLM choices.

In summary, the key contributions of this work include:

\begin{itemize}
    \item We argue that an effective financial deep research agent must systematically integrate authoritative data sources, domain-specific analytical tools, and task-oriented sub-agents to deliver professional financial analysis.
    
    \item Building upon this principle, we propose \textbf{\framework{}},  an end-to-end financial deep research framework that unifies heterogeneous evidence compilation across curated financial data sources, eight domain-specific analytical skills, and two dedicated financial sub-agents into a cohesive analysis pipeline.

    \item We introduce \textbf{TabAgent} and \textbf{HeteroAgent}, two lightweight sub-agents designed specifically for cross-market financial table understanding and cross-modality, heterogeneous financial data interpretation, respectively.

     \item We conduct extensive experiments on three representative financial analysis benchmarks, demonstrating substantial improvements over competitive baselines, with ablation studies further validating the effectiveness of each proposed sub-agent.

\end{itemize}

\begin{figure*}[t]
\centering
\includegraphics[width=\textwidth]{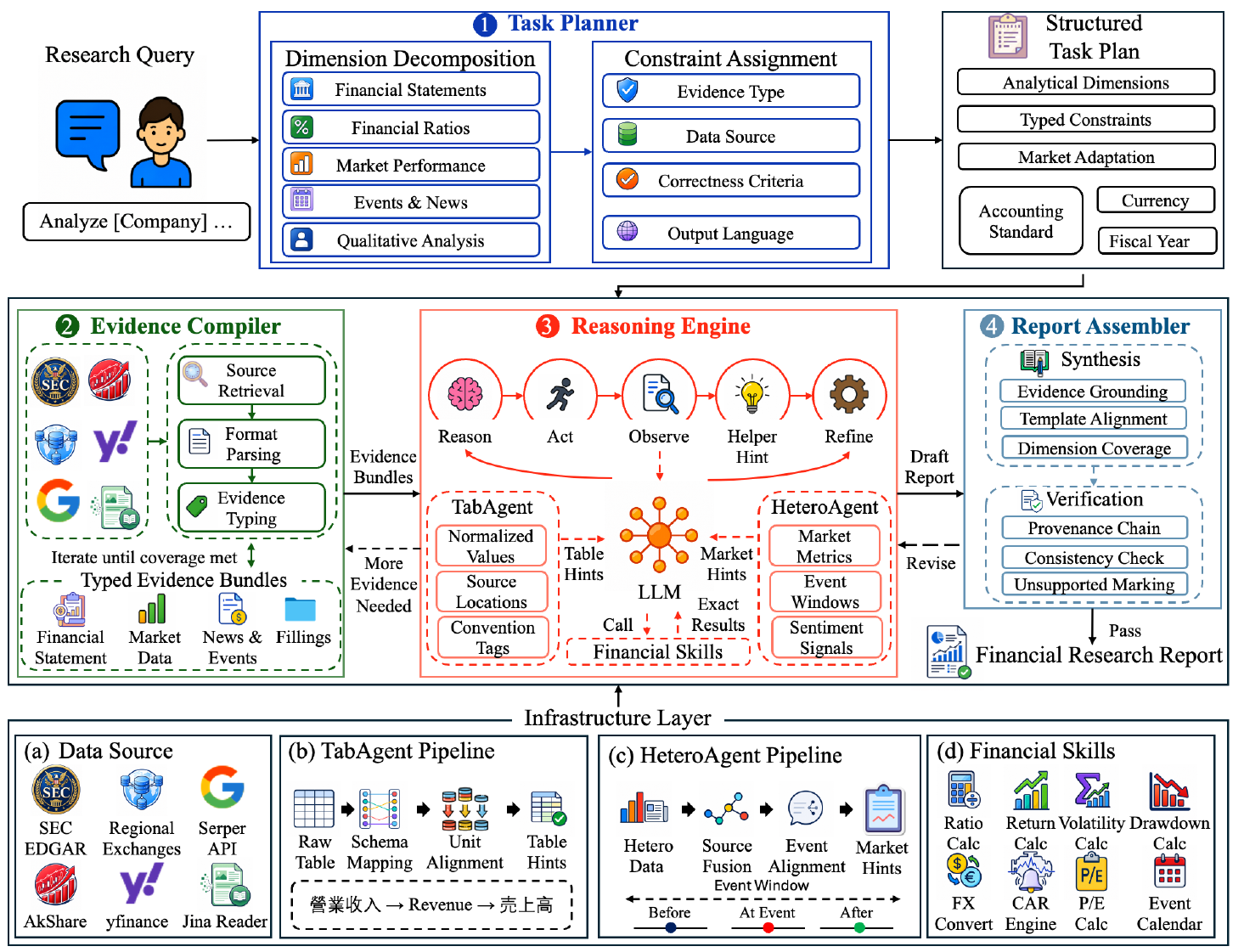}
\caption{Overview of the \framework{} framework.}
\label{fig:framework}
\vspace{-10pt}
\end{figure*}

\section{\framework{} Framework}
\label{sec:method}

\subsection{Problem Formulation}
\label{sec:formulation}

Given a research query $q$ and a target company $\xi$, the task is to produce a structured financial analysis report $\mathcal{R} = \{(c_i, v_i, p_i)\}_{i=1}^{N}$, where each analytical item $c_i$ is assigned a value $v_i$ and a provenance witness $p_i$ tracing $v_i$ to its supporting evidence.
Each item is associated with a correctness predicate $f_i: (v_i, p_i) \rightarrow \{0,1\}$ that determines acceptability.
The objective is:
\begin{equation}
\label{eq:objective}
    \mathcal{R}^* = \arg\max_{\mathcal{R}} \; P(\mathcal{R} \mid q, \xi) \quad \text{s.t.} \quad f_i(v_i, p_i) = 1, \; \forall \, i \in [N]
\end{equation}
The predicates $\{f_i\}_{i=1}^{N}$ fall into structurally incompatible types: exact numerical match, formula-level verification, temporal alignment, and source attribution.
Since the inductive biases required for numerical computation conflict with those for temporal reasoning or source grounding, a single generation pass cannot reliably satisfy all predicate types at once, which motivates a decomposition into typed stages, each equipped with operators suited to the predicate types it handles.

\subsection{Overview}
\label{sec:overview}

\framework{} addresses Eq.~\eqref{eq:objective} by decomposing the generation process into four typed stages, as illustrated in Figure~\ref{fig:framework}.
The data flow is:
\begin{equation}
\label{eq:dataflow}
    (q, \xi) \xrightarrow{\;\text{Planner}\;} \mathcal{P} \xrightarrow{\;\text{Compiler}\;} \mathcal{E} \xrightarrow{\;\text{Engine}\;} \mathcal{R}' \xrightarrow{\;\text{Assembler}\;} \mathcal{R}
\end{equation}
The \emph{Task Planner} decomposes the query into a structured task plan $\mathcal{P}$ that assigns each item $c_i$ a typed constraint specifying its predicate type, required evidence, and target data source.
The \emph{Evidence Compiler} iteratively retrieves and normalises heterogeneous evidence into typed bundles $\mathcal{E}$ under the guidance of $\mathcal{P}$.
The \emph{Reasoning Engine} processes each item under the joint constraints of $\mathcal{P}$ and $\mathcal{E}$, assisted by two role-specialised sub-agents (\textsc{TabAgent} and \textsc{HeteroAgent}) and a suite of Financial Skills.
The \emph{Report Assembler} synthesises the draft report, verifies structural invariants, and triggers revision when verification fails, producing the report $\mathcal{R}$.

\subsection{Task Planner}
\label{sec:planner}

The Task Planner analyses the research query and the target company's market context, then decomposes the task into $K$ dimensions $\{D_1, \ldots, D_K\}$ (financial statements, financial ratios, market performance, events and news, qualitative assessment).
Each dimension groups related analytical items, though items within the same dimension may require different predicate types (e.g., financial ratios involve both exact numerical match and formula-level verification).
For every item $c_i$, the planner assigns a typed constraint $\tau_i$:
\begin{equation}
\label{eq:plan}
    \mathcal{P} = \{\tau_i\}_{i=1}^{N}, \quad \tau_i = (d_i,\, e_i,\, s_i,\, f_i,\, l_i)
\end{equation}
where $d_i \in \{D_k\}$ is the analytical dimension, $e_i$ is the required evidence type, $s_i$ is the target data source, $f_i$ is the correctness predicate, and $l_i$ is the output language.
The plan generation is conditioned on the query and company context:
\begin{equation}
\label{eq:plangen}
    P_\theta(\mathcal{P} \mid q, \xi) = \prod_{i=1}^{N} P_\theta(\tau_i \mid q, \xi, \tau_{<i})
\end{equation}
The planner further performs market adaptation, selecting the applicable accounting standard (US~GAAP, IFRS, HKFRS, PRC~GAAP, or SG-FRS), reporting currency, and fiscal year cycle based on $\xi$.
Once produced, $\mathcal{P}$ acts as a typed contract: downstream components operate within the constraint space defined by $\mathcal{P}$ and cannot produce output that violates it.

\subsection{Evidence Compiler}
\label{sec:compiler}
The Evidence Compiler translates the evidence requirements declared in $\mathcal{P}$ into typed evidence bundles through a three-stage pipeline that iterates until sufficient coverage is achieved.
In the \emph{source retrieval} stage, the compiler constructs dimension-specific retrieval strategies from the evidence types and data source targets specified in $\mathcal{P}$, accessing regulatory filings from SEC~EDGAR and regional exchange databases, market data from yfinance and AkShare, and web content via the Serper search API and Jina Reader, with each round guided by an LLM that identifies uncovered dimensions and generates targeted queries.
In the \emph{format parsing} stage, raw documents in heterogeneous formats (PDF, HTML, XBRL) are parsed into structured content, with financial tables, narrative sections, and metadata separated to preserve their original structure.
In the \emph{evidence typing} stage, each piece of evidence is tagged with its semantic category (financial statement, market data, news and events, or filing metadata) and the analytical dimension it serves, yielding a set of typed evidence bundles:
\begin{equation}
\label{eq:bundle}
    \mathcal{E} = \{b_j\}_{j=1}^{M}, \quad b_j = (\text{content}_j,\, \text{type}_j,\, \text{source}_j,\, d_j,\, t_j,\, u_j,\, \rho_j)
\end{equation}
where $t_j$ denotes the reporting period or timestamp, $u_j$ the unit or currency convention, and $\rho_j$ the provenance metadata (filing identifier, page number, table position, and retrieval URL).
Coupling content with these fields gives downstream provenance verification and source attribution concrete anchors rather than requiring post-hoc reconstruction, and lets heterogeneous evidence be consumed through a unified schema instead of being flattened into an undifferentiated context.
After each iteration the compiler measures the fraction of dimensions in $\mathcal{P}$ whose required evidence types are present in $\mathcal{E}$:
\begin{equation}
\label{eq:coverage}
    \textsc{Cov}(\mathcal{E}, \mathcal{P}) = \frac{|\{d_k \in \mathcal{D} \mid \exists\, b_j \in \mathcal{E},\, d_j = d_k \}|}{|\mathcal{D}|}
\end{equation}
If coverage falls below a threshold $\eta$, a new retrieval round targets the uncovered dimensions, and the loop continues until $\textsc{Cov}(\mathcal{E}, \mathcal{P}) \geq \eta$ or a maximum number of iterations is reached, so that the Reasoning Engine receives pre-classified, pre-normalised evidence with sufficient dimensional coverage.

\subsection{Reasoning Engine}
\label{sec:engine}
The Reasoning Engine is the core component of \framework{}, responsible for filling each analytical item under the joint constraints of $\mathcal{P}$ and $\mathcal{E}$.

\paragraph{Typed ReAct Loop.}
We extend the ReAct framework~\cite{yao2023react} with typed action-space constraints, structured helper hints from role-specialised sub-agents, and callable Financial Skills.
At step $t$ of the loop for dimension $D_k$, the backbone LLM generates an action $a_t$ from a typed action space $\mathcal{A}(\tau_i)$ determined by the constraint $\tau_i$, observes the result $o_t$, receives helper hints $\mathcal{H}_t$ from sub-agents, and updates its reasoning state:
\begin{equation}
\label{eq:react}
    P_\theta(v_i, p_i \mid c_i, \mathcal{P}, \mathcal{E}) = \prod_{t=1}^{T_i} \underbrace{P_\theta(a_t \mid \mathbf{s}_t, \tau_i)}_{\text{typed action}} \cdot \underbrace{P_\theta(\mathbf{s}_{t+1} \mid \mathbf{s}_t, a_t, o_t, \mathcal{H}_t)}_{\text{state update with hints}}
\end{equation}
where $\mathbf{s}_t$ encodes the current reasoning trace, filled values, and accumulated evidence.
$\mathcal{A}(\tau_i)$ restricts the available actions to the predicate type of $c_i$, so that items requiring structured tabular evidence invoke \textsc{TabAgent} hints, items requiring cross-modality evidence invoke \textsc{HeteroAgent} hints, and items requiring exact numerical results invoke Financial Skills, narrowing the generation space at plan time rather than relying on the model to self-regulate at generation time.

\paragraph{TabAgent.}
Financial tables vary widely in language, layout, and accounting convention, so \textsc{TabAgent} standardises raw tabular evidence into a form directly consumable by the backbone LLM through a three-stage pipeline, $\mathcal{H}^{\text{tab}} = \textsc{TabAgent}(\mathcal{E}_{\text{table}},\, \tau_i;\, \theta_{\text{SA1}})$.
In the \emph{schema mapping} stage, heterogeneous line-item names are aligned to a unified canonical schema: \begin{CJK*}{UTF8}{bsmi}``營業收入''\end{CJK*}, ``Revenue'', and \begin{CJK*}{UTF8}{min}``売上高''\end{CJK*} map to the same canonical field regardless of disclosure language, while the accounting standard specified in $\tau_i$ lets the agent separate items that look alike but differ in accounting semantics, such as operating profit and pre-tax profit under PRC~GAAP.
In the \emph{unit alignment} stage, values are normalised to consistent measurement units, sign conventions, and currency denominations.
In the \emph{hint construction} stage, the normalised fields are assembled into table hints $\mathcal{H}^{\text{tab}}$ carrying \emph{normalised values} aligned to the target schema, \emph{source locations} giving the exact filing and page of each value, and \emph{convention tags} recording the applicable accounting standard and sign convention.
The schema mapping rules can be extended to new reporting standards or document types without modifying the core Reasoning Engine.

\paragraph{HeteroAgent.}
When an analytical task requires joint reasoning over structured numerical data, semi-structured documents, and unstructured text, each with its own temporal granularity and reliability profile, a single-pass language model struggles to maintain type consistency across modalities.
\textsc{HeteroAgent} therefore acts as a modular evidence proposer, $\mathcal{H}^{\text{het}} = \textsc{HeteroAgent}(\mathcal{E}_{\text{hetero}},\, \tau_i;\, \theta_{\text{SA2}})$, operating in two general-purpose stages.
The \emph{source fusion} stage places sources of heterogeneous format and temporal granularity, such as daily stock prices, irregularly timestamped news articles, and quarterly disclosures, onto a unified timeline through three operators: \emph{temporal alignment}, which synchronises evidence streams of different frequencies onto a common time axis; \emph{conflict resolution}, which adjudicates discrepancies between sources by prioritising the more authoritative one; and \emph{missing-field completion}, which fills gaps in one source from complementary information in others.
The \emph{event alignment} stage maps detected events such as earnings releases, regulatory actions, and product announcements to exchange trading dates, and constructs event windows $[t_e - w,\, t_e + w]$ for measuring responses before, at, and after each event.
Both stages are source-agnostic and accommodate any evidence modality carrying a timestamp and a typed value.
The output $\mathcal{H}^{\text{het}}$ accordingly consists of \emph{market metrics} (returns, volatility, drawdown over specified windows), \emph{event windows} with precisely aligned temporal boundaries, and \emph{sentiment signals} derived from news semantic analysis.

Both sub-agents are strictly evidence proposers: they supply typed intermediate evidence $\mathcal{H}$ but render no final values, leaving the backbone LLM full authority over reasoning and synthesis using the hints alongside $\mathcal{E}$ and its own chain-of-thought.
New evidence types can therefore be supported by adding or extending sub-agents without modifying the core Reasoning Engine, and the ablations in Section~\ref{sec:experiments} confirm that this role separation outperforms configurations where the same sub-agent model answers directly.

\paragraph{Financial Skills.}
For items whose predicate $f_i$ demands exact numerical match, the Reasoning Engine invokes Financial Skills, a suite of executable functions $g_k: \mathcal{E} \rightarrow v_i$ that compute outside the language model's generation path, comprising Ratio, Return, Volatility, and Drawdown Calculators, an FX Converter, a CAR Engine, a P/E Calculator, and Calendar utilities for trading-day resolution.
The backbone LLM calls these skills during the Act step and receives exact results, eliminating the numerical errors that arise when computation is performed within free-form text generation.

\subsection{Report Assembler}
\label{sec:assembler}
The Report Assembler turns the draft report $\mathcal{R}'$ from the Reasoning Engine into the final report $\mathcal{R}$ through a synthesis-then-verification process.
Synthesis applies \emph{evidence grounding}, which traces every claim to specific evidence in $\mathcal{E}$; \emph{template alignment}, which conforms the output structure to the analytical template defined by $\mathcal{P}$; and \emph{dimension coverage}, which checks that all analytical dimensions have been addressed.
Verification then applies \emph{provenance chain} validation, ensuring that every item $(c_i, v_i, p_i)$ carries a complete evidence trail satisfying $f_i$; \emph{consistency check}, verifying numerical invariants across related items such as component figures summing to reported totals; and \emph{unsupported marking}, which declares items whose evidence cannot satisfy $f_i$ as unsupported rather than filling them with plausible but unverifiable values.
Items failing provenance or consistency checks are returned to the Reasoning Engine for targeted revision over at most two rounds, after which any item still unverified is marked unsupported and the report is output.

\section{Experiments}
\label{sec:experiments}

\subsection{Experimental Setup}
\label{sec:setup}
 
\subsubsection{Benchmarks.}
We evaluate \framework{} on three complementary financial research
benchmarks. FinDeepResearch~\cite{zhu2025findeepresearch}, our primary
benchmark, evaluates deep research agents on full-length corporate
analysis reports for 64 listed companies across 8 markets and 4
languages, with 15{,}808 fine-grained grading items; its HisRubric
framework covers 6 report sections and 4 progressive capability levels:
\emph{Recognition}, \emph{Calculation}, \emph{Abstraction}, and
\emph{Interpretation}. Finance Agent
Benchmark~\cite{bigeard2025finance} evaluates agentic question
answering over SEC filings, covering quantitative retrieval,
qualitative retrieval, numerical reasoning, and cross-document
synthesis; we use its public v1.1 split of 50 questions.
FinTMMBench-Web is our Web-retrieval variant of
FinTMMBench~\cite{zhu2025fintmmbench}. It retains the original 5{,}676
questions over financial tables, news, daily stock prices, and
technical charts from NASDAQ-100 companies, but removes access to the
benchmark's fixed local corpus, so that methods must retrieve the
required evidence from the Web. Together, the three benchmarks cover
full-report analysis, filing-grounded question answering, and temporal
multimodal reasoning under Web retrieval.

\begin{figure}[t]
\centering
\includegraphics[width=0.5\textwidth]{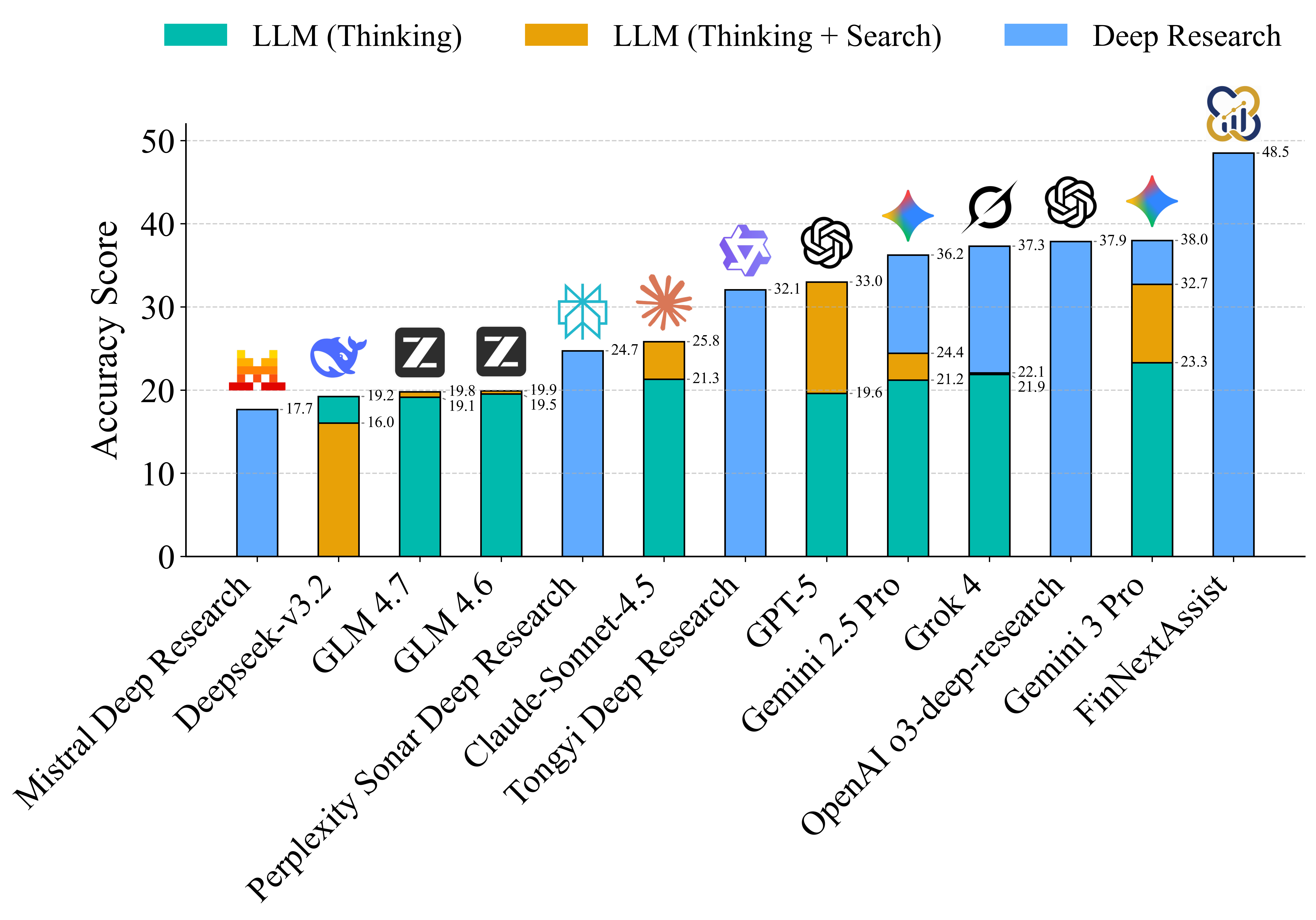}
\caption{Information Precision of all evaluated methods on the FinDeepResearch benchmark.}
\label{fig:main}
\vspace{-0.3cm}
\end{figure}

\subsubsection{Baselines.}
For FinDeepResearch, we compare \framework{} with 23 baselines under
three settings. The \emph{Thinking}~(T) and \emph{Thinking+Search}~(T+S)
settings use the same eight models---OpenAI GPT-5, Claude-Sonnet-4.5,
Gemini~2.5~Pro, Gemini~3~Pro, DeepSeek-v3.2, Grok~4, GLM-4.6, and
GLM-4.7~\cite{gpt5,claudesonnet45,gemini25pro,gemini3pro,deepseekv32,grok4DR,glm46,glm47}---with
the latter additionally enabling search. The \emph{Deep Research}~(DR)
setting includes OpenAI o3-deep-research, Gemini~2.5~Pro Deep Research,
Gemini~3~Pro Deep Research, Grok~4 DeepSearch, Perplexity Sonar Deep
Research, Tongyi Deep Research, and Mistral Deep
Research~\cite{openai2025dr,google2024geminidr,gemini3pro,grok4DR,perplexitydeepresearch,tongyiDR,mistralDR}.
For Finance Agent Benchmark, we select five baselines from the same
model pool: Claude-Sonnet-4.5, OpenAI GPT-5, Gemini~2.5~Pro,
DeepSeek-v3.2, and Grok~4. For FinTMMBench-Web, we evaluate
DeepSeek-v3.2, Gemini~2.5~Pro, and Claude-Sonnet-4.5 under the Web
retrieval setting, and retain the published results of GPT-4o-mini,
BM25, BGE-Text, BGE-Visual, and TMMHybridRAG on the original
FinTMMBench as reference points; the published baselines use the
benchmark's fixed local corpus, whereas the three new baselines and
\framework{} retrieve evidence from the Web.
 
\subsubsection{Implementation and Evaluation.}
The Reasoning Engine uses Claude-Sonnet-4.5~\cite{claudesonnet45} as
its backbone, with a maximum output length of 8{,}192 tokens, at most
10 ReAct rounds per analytical dimension, and a total token budget of
200{,}000. Both \textsc{TabAgent} and \textsc{HeteroAgent} are built on
Qwen-2.5-7B-Instruct~\cite{qwen2024qwen25} through the OpenRouter API.
The Evidence Compiler retrieves regulatory filings from SEC~EDGAR and
regional exchange databases, market data from yfinance and AkShare, and
web content through the Serper search API and Jina Reader. We follow
the official format and evaluation protocol of each benchmark, except
for the retrieval setting of FinTMMBench-Web, and apply no
post-processing or manual correction before evaluation.
FinDeepResearch reports use its prescribed Markdown format with the
output language determined by the Task Planner from the market context;
they are graded by its official evaluation service, which scores 247
items per company against expert-defined rubrics with a total full mark
of 350, and we report Information Precision, defined as the percentage of full
marks achieved, at the overall, section, and capability levels.
Finance Agent Benchmark uses its official harness and tool environment;
we report accuracy on the public v1.1 split, as the full private test
set is not publicly available. For FinTMMBench-Web, methods cannot
access the original benchmark corpus and must retrieve evidence through
Web search; for \framework{}, the retrieved tables, news, price records, and visual evidence pass through the same analysis pipeline, with \textsc{HeteroAgent} aligning them onto a common timeline and visual evidence supplied directly to the backbone. We retain the original questions and report Exact Match (EM), F1, Accuracy, and LLM-judge Accuracy.

\subsection{Main Results}
\label{sec:main_results}
 
\subsubsection{Results on FinDeepResearch.}

Figure~\ref{fig:main} presents the Information Precision of all 24 evaluated methods.
\framework{} achieves a new state of the art of 48.5, outperforming the strongest baseline, Gemini~3~Pro Deep Research (38.0), by 10.5 points.
The baselines progress across paradigm groups, averaging 20.6 for Thinking-only, 24.2 for Thinking+Search, and 32.0 for Deep Research, which confirms that external information acquisition is necessary for financial analysis while also exposing its limits.
Across the eight paired T versus T+S comparisons, search augmentation yields gains ranging from $+$13.4 for GPT-5 to $-$3.2 for DeepSeek-v3.2, with four of the eight models gaining under one point, indicating that retrieving additional evidence without structuring it can introduce as much noise as signal.
The same pattern holds within the Deep Research group, where performance ranges from 17.7 to 38.0 and two Thinking+Search methods surpass the weaker DR systems, so the quality of evidence integration matters more than the research paradigm itself.
\framework{} addresses these limitations by retrieving from authoritative structured sources rather than general web search, by handling numerical computation with executable skills outside the generation path, and by supplying typed evidence hints from role-specialised sub-agents instead of requiring a single model to resolve all evidence types from an undifferentiated context.

\begin{figure*}[t]
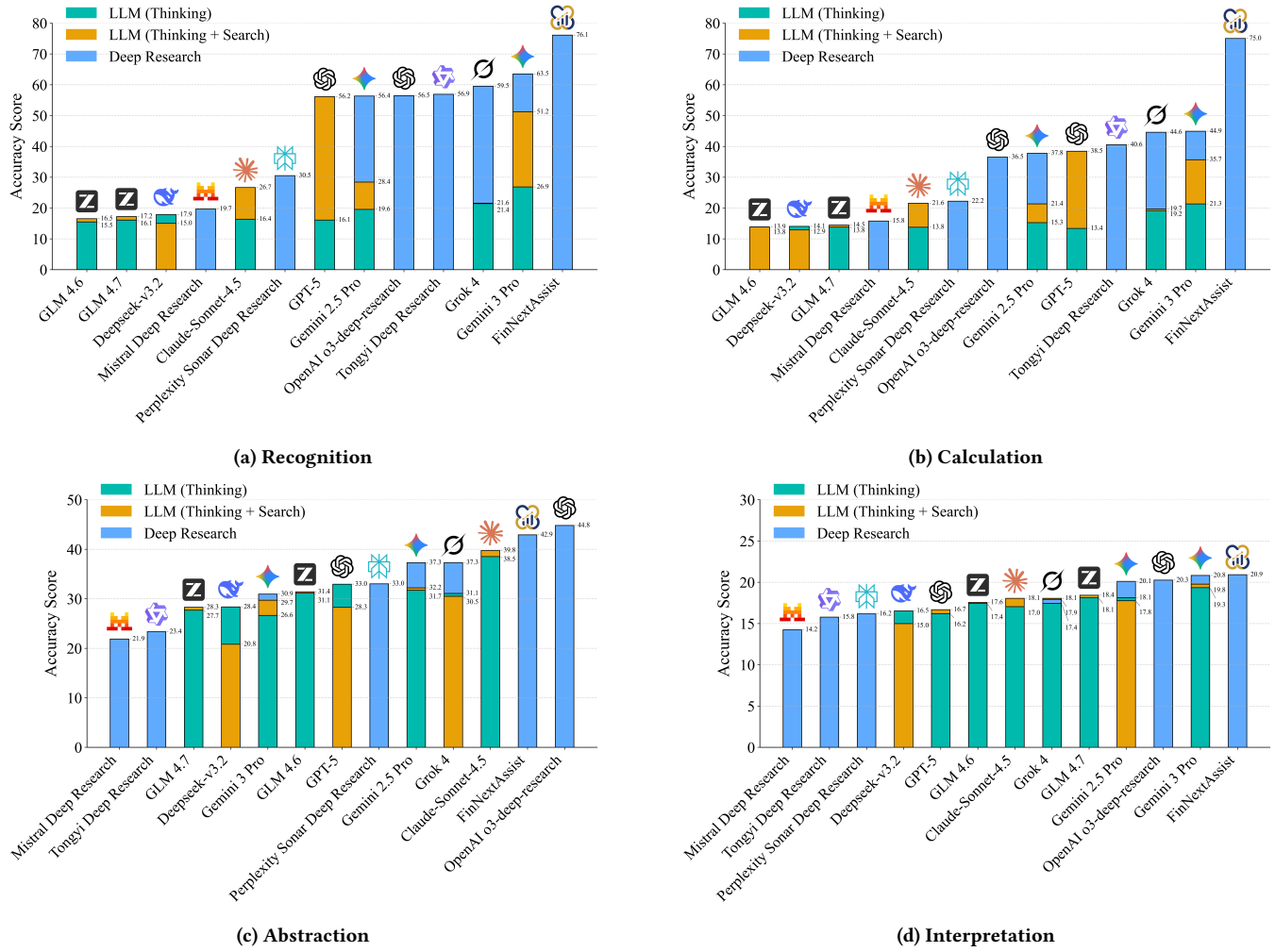

  \centering
  \begin{subfigure}[b]{0.47\textwidth}
    \centering
    
    \includegraphics[width=\linewidth]{fig/Recognition.png}
    \caption{Recognition}
    \label{fig:recognition}
  \end{subfigure}
  \hfill
  \begin{subfigure}[b]{0.47\textwidth}
    \centering
    \includegraphics[width=\linewidth]{fig/Calculation.png}
    \caption{Calculation}
    \label{fig:calculation}
  \end{subfigure}
  \begin{subfigure}[b]{0.47\textwidth}
    \centering
    \includegraphics[width=\linewidth]{fig/Abstraction.png}
    \caption{Abstraction}
    \label{fig:abstraction}
  \end{subfigure}
  \hfill
  \begin{subfigure}[b]{0.47\textwidth}
    \centering
\includegraphics[width=\linewidth]{fig/Interpretation.png}
    \caption{Interpretation}
    \label{fig:interpretation}
  \end{subfigure}
  \caption{Performance analysis across four different capabilities.}
  \label{fig:four-tasks}
   \vspace{-0.2cm}
\end{figure*}

\begin{table}[t]
\centering
\caption{Performance analysis on Finance Agent Benchmark.}
\label{tab:fab}
\renewcommand{\arraystretch}{1.1}
\resizebox{\columnwidth}{!}{
\begin{tabular}{lcc}
\toprule
\textbf{Method} & \textbf{Accuracy (\%)} & \textbf{$\Delta$ vs.\ \framework{}} \\
\midrule
DeepSeek-v3.2 & 6.18 & $-$71.21 \\
Gemini 2.5 Pro & 44.53 & $-$32.86 \\
Grok 4 & 51.85 & $-$25.54 \\
OpenAI GPT-5 & 53.02 & $-$24.37 \\
Claude-Sonnet-4.5 & \underline{56.71} & $-$20.68 \\
\midrule
\rowcolor{blue!6} \bf \framework{} & \textbf{77.39} & -- \\
\bottomrule
\end{tabular}
}
\vspace{-0.2cm}
\end{table}

\begin{table}[t]
\setlength{\abovecaptionskip}{2pt}
\setlength{\belowcaptionskip}{0pt}
\centering
\caption{Performance on FinTMMBench-Web.\textbf{Bold} marks the best result overall; \underline{underline} marks the best Web-retrieval baseline.}
\label{tab:fintmm-main}
\renewcommand{\arraystretch}{1.05}
\setlength{\tabcolsep}{3.2pt}
\resizebox{\columnwidth}{!}{%
\begin{tabular}{lccccc}
\toprule
\textbf{Method} &
\makecell{\textbf{Retrieval}\\\textbf{Corpus}} &
\textbf{EM} &
\textbf{F1} &
\textbf{Acc.} &
\textbf{LLM Acc.} \\
\midrule
GPT-4o-mini       & Local & 4.51  & 5.89  & 6.68  & 6.25  \\
BM25              & Local & 10.85 & 20.89 & 15.89 & 11.97 \\
BGE-Text          & Local & 17.11 & 27.36 & 23.41 & 18.50 \\
BGE-Visual        & Local & 14.51 & 25.45 & 21.89 & 16.04 \\
TMMHybridRAG      & Local & 19.12 & 31.41 & 26.56 & 21.53 \\
\midrule
DeepSeek-v3.2     & Web & \underline{6.84} & \underline{10.78}
                  & 7.80 & \underline{12.67} \\
Gemini~2.5~Pro    & Web & 5.53 & 7.59 & 6.22 & 8.54 \\
Claude-Sonnet-4.5 & Web & 3.49 & 9.95 & \underline{13.30} & 10.59 \\
\midrule
\rowcolor{blue!6}
\textbf{\framework{}} & \textbf{Web} & \textbf{32.71}
& \textbf{52.04} & \textbf{47.21} & \textbf{42.69} \\
\bottomrule
\end{tabular}%
}
\vspace{-0.6cm}
\end{table}

\subsubsection{Results on Finance Agent Benchmark.}

To verify that the improvements generalize beyond FinDeepResearch, we conduct an external validation on the public v1.1 split of the Finance Agent Benchmark~\cite{bigeard2025finance}, which evaluates
tool-augmented agents on real-world financial research tasks requiring SEC filing retrieval, document parsing, and numerical extraction. As shown in Table~\ref{tab:fab}, \framework{} achieves 77.39\%
accuracy, outperforming the strongest baseline, Claude-Sonnet-4.5 (56.71\%), by 20.68 points. GPT-5 (53.02\%), Grok~4 (51.85\%), and Gemini~2.5~Pro (44.53\%) trail by wider margins, while DeepSeek-v3.2 scores 6.18\%.
 
The comparison between Claude-Sonnet-4.5 as a direct baseline (56.71\%) and as the backbone model within \framework{} (77.39\%) is informative. The same underlying model gains 20.68 points when augmented with the structured evidence pipeline, confirming that the improvement is attributable to the framework design rather than model capacity. This gap is larger than the 10.5-point improvement observed on FinDeepResearch, which may reflect the fact that the Finance Agent
Benchmark's question-level tasks place greater emphasis on precise evidence retrieval and numerical extraction from individual filings, where structured data access and deterministic computation have a more direct impact.

\subsubsection{Results on FinTMMBench-Web.}
Table~\ref{tab:fintmm-main} reports results under the Local and Web retrieval settings. Among the Web baselines, DeepSeek-v3.2 is strongest on EM, F1, and LLM-judge Accuracy and Claude-Sonnet-4.5 on Accuracy,
yet the best Web baseline result on any metric is 13.30, indicating that Web access alone is insufficient for this task. \framework{} reaches 32.71 EM, 52.04 F1, 47.21 Accuracy, and 42.69 LLM-judge Accuracy, exceeding the best Web baseline on each metric by 25.87,
41.26, 33.91, and 30.02 points. It also scores above the local-corpus reference points, including TMMHybridRAG (19.12 EM, 31.41 F1), although those results are obtained under a different retrieval setting and are
therefore not directly comparable. As with the local-corpus methods, LLM-judge Accuracy falls below Accuracy for \framework{}.
 
The relation between the surface-form metric and the content metrics is informative: F1 and Accuracy exceed EM by 19.33 and 14.50 points respectively, so the answers often carry the correct content even when
they do not match the reference string exactly. This is consistent with how the answer is produced: the typed evidence interface constrains what an answer must be grounded in, not the form in which it is expressed.
 
The three benchmarks differ substantially in task format and evaluation protocol. FinDeepResearch grades full-length reports on 247 items per company, Finance Agent Benchmark targets question answering over
individual SEC filings, and FinTMMBench-Web requires temporal multimodal reasoning under Web retrieval. \framework{} leads under all three, indicating that the typed evidence interface transfers across task granularities and evidence types rather than being tuned to a single task format.

\subsection{In-depth Analysis}

\begin{table}[t]
\centering
\caption{Performance analysis across six sections. Values denote normalized accuracy scores. The best and second-best methods are indicated with \textbf{bold} and \underline{underline}.}
\label{tab:sections}
\renewcommand{\arraystretch}{0.9}
\setlength{\tabcolsep}{0pt}
\begin{tabular*}{\columnwidth}{@{}L{0.45\columnwidth}@{\extracolsep{\fill}}rrrrrr@{}}
\toprule
\textbf{Method} & \textbf{S1} & \textbf{S2} & \textbf{S3} & \textbf{S4} & \textbf{S5} & \textbf{S6} \\

\rowcolor{blue!6} \multicolumn{7}{c}{\textit{\textbf{LLM (Thinking)}}} \\
Gemini 2.5 Pro (T) & 39.1 & 15.2 & 32.0 & 37.3 & 44.0 & 14.5 \\
DeepSeek-v3.2 (T) & 37.4 & 14.6 & 28.8 & 33.6 & 36.8 & 12.8 \\
Claude-Sonnet-4.5 (T) & 42.0 & 13.0 & 36.1 & 47.7 & 47.3 & 13.3 \\
Grok 4 (T) & 42.8 & 19.2 & 30.1 & 36.9 & 37.7 & 13.9 \\
OpenAI GPT-5 (T) & 41.7 & 12.8 & 29.5 & 40.5 & 41.8 & 12.6 \\
Gemini 3 Pro Preview (T) & 37.3 & 23.0 & 29.1 & 30.3 & 40.0 & 15.8 \\
GLM 4.6 (T) & 38.3 & 12.5 & 30.4 & 37.7 & 40.4 & 13.6 \\
GLM 4.7 (T) & 36.6 & 13.1 & 28.1 & 33.6 & 35.8 & 14.3 \\

\rowcolor{blue!6} \multicolumn{7}{c}{\textit{\textbf{LLM (Thinking + Search)}}} \\
Gemini 2.5 Pro (T+S) & 40.3 & 24.2 & 33.9 & 36.8 & 43.3 & 14.5 \\
DeepSeek-v3.2 (T+S) & 29.3 & 12.7 & 20.4 & 24.9 & 28.8 & 12.2 \\
Claude-Sonnet-4.5 (T+S) & 41.3 & 23.3 & 41.0 & \underline{48.7} & 49.5 & 14.4 \\
Grok 4 (T+S) & 41.4 & 19.3 & 30.4 & 34.9 & 37.5 & 14.7 \\
OpenAI GPT-5 (T+S) & 38.2 & 53.9 & 32.9 & 33.0 & 37.7 & 13.2 \\
Gemini 3 Pro Preview (T+S) & 39.3 & 47.6 & 33.1 & 32.9 & 43.2 & 16.6 \\
GLM 4.6 (T+S) & 36.8 & 13.2 & 31.7 & 37.4 & 41.2 & 13.8 \\
GLM 4.7 (T+S) & 35.8 & 14.1 & 30.0 & 34.1 & 36.9 & 14.6 \\

\rowcolor{blue!6} \multicolumn{7}{c}{\textit{\textbf{Deep Research}}} \\
Perplexity Deep Research & 39.7 & 26.9 & 34.8 & 40.5 & 41.0 & 12.9 \\
Mistral Deep Research & 35.6 & 16.4 & 20.6 & 26.2 & 29.3 & 11.9 \\
Tongyi Deep Research & 36.4 & 53.4 & 26.1 & 28.9 & 35.8 & 14.0 \\
Gemini 2.5 Pro Deep Research & 47.3 & 52.4 & 38.1 & 39.2 & \underline{49.9} & 17.0 \\
Grok 4 DeepSearch & \textbf{56.1} & 57.2 & 38.1 & 41.4 & 41.6 & 15.5 \\
OpenAI o3-deep-research & \underline{49.9} & 52.0 & \underline{45.4} & \textbf{52.9} & \textbf{56.5} & 16.4 \\
Gemini 3 Pro Deep Research & 42.2 & \underline{60.7} & 34.1 & 34.1 & 44.8 & \underline{17.8} \\

\rowcolor{blue!6} \multicolumn{7}{c}{\textit{\textbf{Ours}}} \\
\bf \framework{} & 43.2 & \textbf{77.0} & \textbf{53.4} & 44.8 & 44.4 & \textbf{24.6} \\
\bottomrule
\end{tabular*}
\vspace{-0.2cm}
\end{table}

\begin{table}[t]
\centering
\caption{Performance analysis across eight financial markets. Values denote normalized accuracy scores. The best and second-best methods are indicated with \textbf{bold} and \underline{underline}.}
\label{tab:markets}
\renewcommand{\arraystretch}{0.9}
\setlength{\tabcolsep}{0pt}
\begin{tabular*}{\columnwidth}{@{}L{0.45\columnwidth}@{\extracolsep{\fill}}rrrrrrrr@{}}
\toprule
\multirow{2}{*}{\textbf{Method}}
& \includegraphics[width=0.36cm]{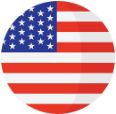}
& \includegraphics[width=0.36cm]{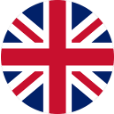}
& \includegraphics[width=0.36cm]{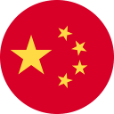}
& \includegraphics[width=0.36cm]{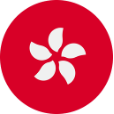}
& \includegraphics[width=0.36cm]{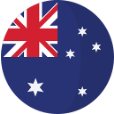}
& \includegraphics[width=0.36cm]{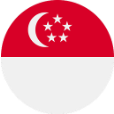}
& \includegraphics[width=0.36cm]{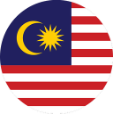}
& \includegraphics[width=0.36cm]{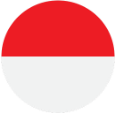} \\
& \textbf{US} & \textbf{UK} & \textbf{CN} & \textbf{HK} & \textbf{AU} & \textbf{SG} & \textbf{MY} & \textbf{ID} \\

\rowcolor{blue!6} \multicolumn{9}{c}{\textit{\textbf{LLM (Thinking)}}} \\
Gemini 2.5 Pro (T) & 19.9 & 21.0 & 17.6 & 20.8 & 24.4 & 24.2 & 25.1 & 16.5 \\
DeepSeek-v3.2 (T) & 19.7 & 17.7 & 17.3 & 18.4 & 20.9 & 21.0 & 23.8 & 15.0 \\
Claude-Sonnet-4.5 (T) & 22.2 & 19.9 & 19.1 & 21.7 & 23.0 & 22.7 & 24.7 & 17.0 \\
Grok 4 (T) & 23.2 & 24.0 & 16.9 & 18.4 & 25.8 & 24.3 & 25.0 & 17.4 \\
OpenAI GPT-5 (T) & 18.1 & 18.7 & 16.6 & 17.6 & 22.6 & 23.6 & 23.3 & 16.3 \\
Gemini 3 Pro Preview (T) & 22.4 & 23.4 & 21.3 & 22.1 & 26.4 & 24.9 & 25.9 & 20.1 \\
GLM 4.6 (T) & 18.3 & 17.2 & 18.2 & 18.5 & 20.9 & 22.7 & 24.0 & 16.6 \\
GLM 4.7 (T) & 16.5 & 18.0 & 16.5 & 19.5 & 21.3 & 21.2 & 24.2 & 16.0 \\

\rowcolor{blue!6} \multicolumn{9}{c}{\textit{\textbf{LLM (Thinking + Search)}}} \\
Gemini 2.5 Pro (T+S) & 22.9 & 20.7 & 20.4 & 24.7 & 26.4 & 29.2 & 27.5 & 23.6 \\
DeepSeek-v3.2 (T+S) & 10.9 & 14.9 & 16.8 & 16.5 & 20.4 & 17.7 & 21.0 & 10.0 \\
Claude-Sonnet-4.5 (T+S) & 27.8 & 23.0 & 25.7 & 20.3 & 27.4 & 28.5 & 30.4 & 23.4 \\
Grok 4 (T+S) & 23.7 & 22.4 & 17.8 & 19.4 & 27.2 & 24.6 & 25.0 & 16.4 \\
OpenAI GPT-5 (T+S) & 37.4 & 36.9 & 20.8 & 29.3 & 35.6 & 42.5 & 32.3 & 29.1 \\
Gemini 3 Pro Preview (T+S) & 33.1 & 30.3 & 29.2 & 24.9 & 36.1 & 39.2 & 37.3 & 31.6 \\
GLM 4.6 (T+S) & 16.8 & 16.8 & 19.1 & 20.2 & 21.9 & 24.0 & 23.7 & 16.6 \\
GLM 4.7 (T+S) & 15.8 & 17.3 & 19.2 & 21.4 & 21.7 & 22.9 & 23.9 & 16.2 \\

\rowcolor{blue!6} \multicolumn{9}{c}{\textit{\textbf{Deep Research}}} \\
Perplexity Deep Research & 21.0 & 23.7 & 22.4 & 25.0 & 28.8 & 26.9 & 26.9 & 23.0 \\
Mistral Deep Research & 13.5 & 16.1 & 14.0 & 13.6 & 22.2 & 21.1 & 23.7 & 17.1 \\
Tongyi Deep Research & 32.1 & 27.8 & 27.8 & 29.5 & 36.1 & 35.6 & 37.3 & 30.3 \\
Gemini 2.5 Pro Deep Research & 37.6 & 34.1 & 30.8 & 36.0 & 36.0 & 38.9 & 39.8 & 36.6 \\
Grok 4 DeepSearch & 34.5 & 39.0 & 33.4 & \underline{36.4} & 39.3 & \underline{46.7} & 37.9 & 31.3 \\
OpenAI o3-deep-research & \underline{42.5} & \underline{43.0} & 34.7 & 30.2 & \underline{41.7} & 33.6 & 38.3 & \underline{38.9} \\
Gemini 3 Pro Deep Research & 39.6 & 34.4 & \underline{34.9} & 34.0 & 41.3 & 40.6 & \underline{42.3} & 36.8 \\

\rowcolor{blue!6} \multicolumn{9}{c}{\textit{\textbf{Ours}}} \\
\bf \framework{} & \textbf{51.7} & \textbf{47.1} & \textbf{49.3} & \textbf{48.3} & \textbf{46.6} & \textbf{50.2} & \textbf{49.8} & \textbf{44.9} \\

\bottomrule
\end{tabular*}
\end{table}

\subsubsection{Performance Analysis Across Sections.}
Table~\ref{tab:sections} reports normalized accuracy across the six sections, where \framework{} ranks first in three.
The largest absolute gain appears in S2 (Financial Performance), where \framework{} obtains 77.0 against 60.7 for Gemini~3~Pro Deep Research, a margin of 16.3 points that aligns with the design of the framework, since S2 covers financial statements, key ratios, and operating performance, all of which depend on evidence normalized by TabAgent and on exact computation by executable skills.
\framework{} also gains 6.8 points in S6 (Market Performance), the most heterogeneous section, reaching 24.6 against 17.8, a 38\% relative improvement despite low absolute scores across all methods, which reflects the value of HeteroAgent in aligning events with valid trading dates and connecting market movements to supporting evidence.
In S3 (Business Analysis) it reaches 53.4 against 45.4 from OpenAI o3-deep-research, indicating that typed intermediate evidence supports not only extraction and calculation but also downstream analytical reasoning.
In S1, S4, and S5, \framework{} remains competitive with leading DR systems without ranking first, as these sections rely on descriptive summarisation and qualitative interpretation.
This delineates the applicability boundary of our approach, which excels when correctness depends on structured data, computation, temporal alignment, and heterogeneous evidence integration, and offers less direct advantage where open-ended narrative synthesis dominates.

\subsubsection{Performance Analysis Across Capabilities.}
Figure~\ref{fig:four-tasks} reports performance across the four capability levels.
\framework{} gains most in Recognition and Calculation, scoring 76.1 ($+$12.6 over Gemini~3~Pro Deep Research) and 75.0 ($+$30.1).
The Recognition gain comes from organizing retrieved information into typed evidence, which reduces factual extraction errors before they propagate downstream.
The Calculation gain is the largest single-capability improvement in our experiments: rather than having the language model infer or regenerate numerical values from context, \framework{} offloads ratio, return, and currency operations to executable skills outside the generation path and grounds statement-level computation in TabAgent, treating numerical sub-tasks as verifiable computation problems rather than text generation.
For Abstraction, \framework{} reaches 42.9, slightly below OpenAI o3-deep-research (44.8) but 3.1 points above its own backbone with search enabled (Claude-Sonnet-4.5 T+S, 39.8), and for Interpretation it obtains 20.9, the highest among all methods and 2.8 points above that backbone (18.1), though the low absolute scores throughout indicate that deep financial interpretation remains the least solved capability in the benchmark.
The structured evidence interface therefore delivers substantial gains where correctness can be tied to typed evidence and executable skills, and moderate but consistent improvement on higher-level reasoning.

\subsubsection{Performance Analysis Across Markets.}
Table~\ref{tab:markets} reports normalized accuracy across eight financial markets, where \framework{} ranks first in every one.
The gains are largest in non-English markets, reaching $+$14.4 in China and $+$11.9 in Hong Kong, where disclosures involve Simplified and Traditional Chinese, which indicates that typed evidence organization is most valuable when information must be extracted from multilingual sources and normalized under market-specific conventions.
Gains remain positive throughout the English-language and Southeast Asian markets, from $+$9.2 in the US down to $+$3.5 in Singapore, with the smaller margins concentrated in markets using internationally aligned reporting standards, where the cross-lingual challenge is less pronounced.
A further pattern concerns cross-market consistency: three different DR agents lead across the eight markets and the best baseline score per market ranges from 34.9 to 46.7, whereas \framework{} maintains a much narrower band from 44.9 to 51.7, suggesting that market-aware planning and typed evidence normalization improve robustness across heterogeneous reporting environments.

\subsection{Ablation Study}

\subsubsection{Effect Analysis of each Sub-agent.}
Table~\ref{tab:subagent-ablation} evaluates the contribution of TabAgent and HeteroAgent by selectively removing them.
The complete system achieves 48.5, while removing TabAgent yields 44.2 ($-$4.3), removing HeteroAgent yields 46.0 ($-$2.5), and removing both yields 42.7 ($-$5.8).
The two sub-agents affect different capabilities: removing TabAgent costs 10.0 points in Recognition (76.1$\to$66.1) and 8.6 in Calculation (75.0$\to$66.4), confirming that structured table extraction and cross-lingual schema normalization are central to identifying financial statement values accurately, whereas removing HeteroAgent barely affects Recognition (76.1$\to$75.9) but reduces Calculation by 13.9 points (75.0$\to$61.1), indicating that it improves the evidence conditions for downstream computation rather than direct factual extraction.
The two also act synergistically on Calculation: relative to the no-agent baseline of 55.0, enabling either alone yields 61.1 ($+$6.1) or 66.4 ($+$11.4), summing to 17.5, yet enabling both reaches 75.0 ($+$20.0), a 2.5-point surplus that suggests TabAgent's normalized table fields and HeteroAgent's aligned heterogeneous evidence jointly provide a richer input than either does in isolation.
Abstraction (41.4--42.9) and Interpretation (20.6--20.9) vary only marginally across all four settings, confirming that the sub-agents strengthen evidence acquisition and numerical correctness while higher-level analytical reasoning remains governed by the backbone.

\begin{table}[t]
\centering
\caption{Ablation study of TabAgent and HeteroAgent.}
\label{tab:subagent-ablation}

\normalsize
\renewcommand{\arraystretch}{1.15}
\setlength{\tabcolsep}{1.8pt}

\begin{tabular*}{\columnwidth}{@{\extracolsep{\fill}}cccccccc@{}}
\toprule
\multicolumn{2}{c}{\textbf{Agent Setting}}
& \multicolumn{2}{c}{\textbf{Overall}}
& \multicolumn{4}{c}{\textbf{Capability Score}} \\
\cmidrule(lr){1-2}
\cmidrule(lr){3-4}
\cmidrule(lr){5-8}
\makecell{\textbf{Tab}\\\textbf{Agent}}
& \makecell{\textbf{Hetero}\\\textbf{Agent}}
& \textbf{Score}
& \makecell{\textbf{$\Delta$}\\}
& \textbf{Recog.}
& \textbf{Calc.}
& \textbf{Abstr.}
& \textbf{Interp.} \\
\midrule
\xmark & \xmark & 42.7 & -5.8 & 66.5 & 55.0 & 41.4 & 20.6 \\
\xmark & \cmark & 44.2 & -4.3 & 66.1 & 66.4 & 41.6 & 20.7 \\
\cmark & \xmark & 46.0 & -2.5 & 75.9 & 61.1 & 41.5 & 20.7 \\
\rowcolor{blue!6}
\cmark & \cmark & \textbf{48.5} & -- & \textbf{76.1} & \textbf{75.0} & \textbf{42.9} & \textbf{20.9} \\
\bottomrule
\end{tabular*}
\end{table}

\begin{table}[t]
\centering
\caption{Ablation study of different LLM backbones.}
\label{tab:backbone-ablation}

\normalsize
\renewcommand{\arraystretch}{1.15}
\setlength{\tabcolsep}{2.2pt}

\begin{tabular*}{\columnwidth}{@{\extracolsep{\fill}}lcccccc@{}}
\toprule
\makecell{\textbf{LLM}\\\textbf{Backbone}}
& \multicolumn{2}{c}{\textbf{Overall}}
& \multicolumn{4}{c}{\textbf{Capability Score}} \\
\cmidrule(lr){2-3}
\cmidrule(lr){4-7}
& \textbf{Score}
& \makecell{\textbf{$\Delta$}}
& \textbf{Recog.}
& \textbf{Calc.}
& \textbf{Abstr.}
& \textbf{Interp.} \\
\midrule
GPT-4o-mini & 46.8 & -1.7 & 75.8 & 67.5 & 41.4 & 20.4 \\
DeepSeek-v3.2 & 47.1 & -1.4 & \textbf{76.2} & 68.9 & 41.6 & 20.3 \\
\rowcolor{blue!6}
\makecell[l]{Claude-Sonnet-4.5}
& \textbf{48.5} & -- & 76.1 & \textbf{75.0} & \textbf{42.9} & \textbf{20.9} \\
\bottomrule
\end{tabular*}
\end{table}

\subsubsection{Analysis of Different LLM Backbones.}
Table~\ref{tab:backbone-ablation} evaluates whether performance depends on a specific backbone.
Claude-Sonnet-4.5 scores highest at 48.5, with DeepSeek-v3.2 and GPT-4o-mini at 47.1 and 46.8, and even the weakest configuration exceeds the strongest baseline in Figure~\ref{fig:main} (38.0) by 8.8 points, confirming that the gain is attributable to the framework design rather than a particular backbone.
Recognition is highly stable across backbones, with DeepSeek-v3.2 even leading at 76.2 against 76.1 and 75.8, which suggests that once evidence is retrieved, normalized, and organized into typed intermediate objects, factual recognition becomes largely independent of the backbone.
Calculation is more sensitive, falling from 75.0 to 68.9 and 67.5, because the backbone still selects relevant evidence, invokes the appropriate skills, and integrates computed values even though the executable skills perform the numerical operations themselves.
Abstraction declines modestly from 42.9 to 41.4 and Interpretation stays between 20.3 and 20.9, so \framework{} is robust to backbone replacement while stronger backbones still benefit coordination-intensive capabilities.

\section{Related Work}
\label{sec:related}

\subsection{Deep Research Agents.}
Deep Research agents tackle complex tasks through planning, retrieval, and multi-step reasoning, building on early agentic paradigms such as Reflexion~\cite{shinn2023reflexion} and on environments like WebArena~\cite{zhou2023webarena} and AgentBench~\cite{liu2025agentbenchevaluatingllmsagents} that made agent capabilities systematically measurable.
Recent commercial~\cite{openai2025dr,tongyiDR,mistralDR} and open-source~\cite{jin2025searchr1,song2025r1searcher} systems combine reasoning-capable LLMs with reinforcement learning for greater autonomy, while evaluation has moved from general research quality~\cite{du2025deepresearchbench} to fine-grained rubric-based assessment of domain-specific analysis~\cite{zhu2025findeepresearch}.
Yet most DR agents still hold retrieved materials in a largely unified context, leaving the model to infer evidence boundaries, source reliability, and verification rules implicitly, and diagnostic studies~\cite{zhang2025finder} attribute DR failures more often to evidence integration than to comprehension.
\framework{} instead makes evidence organization explicit, decomposing retrieved materials into typed intermediate objects with predicate-specific constraints so that numerical consistency, temporal alignment, and source attribution are each handled under their own correctness requirement.

\subsection{LLM-based Financial Systems.}
Domain-adapted models such as BloombergGPT~\cite{wu2023bloomberggpt}, FinBERT~\cite{araci2019finbert}, and FinReport~\cite{10.1145/3589335.3648330} established the foundations of financial language understanding.
For analysis and report generation, FinGPT~\cite{yang2023fingpt} enabled event understanding through instruction tuning, MMRepAgent~\cite{LI2026115197} introduced a multimodal agent framework for earnings forecasting, and FinSight~\cite{jin2025finsight} proposed a code-agent architecture with variable memory for end-to-end financial deep research.
In trading, designs have evolved from single agents~\cite{yu2023finmem} to collaborative frameworks, including QuantAgents~\cite{li2025quantagents} for simulated-trading strategy evaluation, HedgeAgents~\cite{li2025hedgeagents} for hedging-based portfolio coordination, and TradingAgents~\cite{xiao2024tradingagents} for multi-role trading simulation.
These systems organize data, tools, and agents for report-level or decision-level generation, but still reason over heterogeneous evidence in a largely unified process, whereas \framework{} separates evidence by correctness requirement and handles it with typed stages, role-specialised components, and deterministic operators.

\section{Conclusion}
We identified three requirements for effective financial deep research, namely authoritative heterogeneous data sources, specialized financial tools and skills, and dedicated sub-agents for domain-specific sub-tasks, and proposed \framework{}, an end-to-end framework that meets them through a four-stage pipeline of Task Planner, Evidence Compiler, Reasoning Engine, and Report Assembler.
Its two lightweight sub-agents, TabAgent for cross-market financial table understanding and HeteroAgent for cross-modality heterogeneous data interpretation, supply typed evidence to the backbone, while eight executable financial skills keep numerical operations off the generation path, so that heterogeneous reasoning requirements are never conflated within a single undifferentiated context.
Experiments on three financial research benchmarks confirm the effectiveness and generalizability of this design.
Two directions remain open: automatically identifying the required sub-agents and skills from query characteristics rather than curating them manually, and end-to-end sub-agent training with self-evolving adaptation, in which sub-agents are fine-tuned on domain-specific feedback and the framework refines its own planning and routing strategies through experience.

\section{Ethical Considerations}
\label{sec:ethics}
The principal risk of automated financial analysis is that unverifiable claims are presented with the same confidence as verified ones, which in a financial setting can propagate into decisions with material consequences. \framework{} mitigates this by construction: every reported item carries a provenance chain to its supporting evidence, numerical results are produced by executable skills rather than free-form generation, and items that the available evidence cannot substantiate are declared unsupported. All data used are publicly available corporate disclosures, market records, and benchmark corpora, accessed under their respective terms of service; the framework retrieves only firm-level financial information and processes no personal data at any stage.

\bibliographystyle{ACM-Reference-Format}
\bibliography{reference}

\end{document}